\documentclass[aps,prl,fleqn,superscriptaddress,floatfix,twocolumn]{revtex4}

\usepackage{graphicx,color}
\usepackage{amsmath,amssymb,amsfonts}

\newcommand{\be}{\begin{equation}}
\newcommand{\ee}{\end{equation}}
\newcommand{\ba}{\begin{eqnarray}}
\newcommand{\ea}{\end{eqnarray}}
\newcommand{\ban}{\begin{eqnarray*}}
\newcommand{\ean}{\end{eqnarray*}}
\newcommand{\nn}{\nonumber}

\begin{document}

\title{Hadron Mass Spectrum from Lattice QCD}
\medskip

\author{Abhijit Majumder}
\affiliation{Department of Physics, The Ohio State University, Columbus, OH 43210,  USA}

\author{Berndt M\"uller}
\affiliation{Department of Physics \& CTMS, Duke University, Durham, NC 27708, USA}

\date{\today}

\begin{abstract}
Finite temperature lattice simulations of quantum chromodynamics (QCD) are sensitive to the hadronic mass spectrum for temperatures below the ``critical'' temperature $T_c \approx 160$ MeV. We show that a recent precision determination of the QCD trace anomaly shows evidence for the existence of a large number of hadron states beyond those known from experiment. The lattice results are well represented by an exponentially growing hadron mass spectrum up to a temperature $T=155$ MeV. Using simple parametrizations we show how one may estimate the total spectral weight in these yet undermined states.
\end{abstract}

\maketitle

Experimental data of multiparticle production in proton-proton collisions led Hagedorn to propose that the spectrum of hadronic states grows exponentially with mass~\cite{Hagedorn:1965st,Hagedorn:1968zz}. Such a spectrum arises naturally in the dual resonance model~\cite{Veneziano:1974dr} and, more generally, in models of quark confinement, such as string models or bag models~\cite{Johnson:1975sg}. The spectrum of experimentally established hadronic states~\cite{Eidelman:2004wy} is compatible with such an exponential mass spectrum up to masses of approximately 1.7 GeV \cite{Broniowski:2004yh}. Most higher-mass hadron states are difficult to identify experimentally because of their increasingly large width and complicated decay properties. Nevertheless, significant efforts are being made to extend the baryon mass spectrum to higher masses \cite{Kamano:2009mm}, and searches for new meson states, including exotic states beyond those predicted by the constituent quark model, are planned at the upgraded Jefferson Laboratory 12 GeV beam facility~\cite{Meyer:2002zr}.

A first-principles prediction of the hadron mass spectrum from quantum chromodynamics by means of lattice QCD simulations would be highly desirable. Unfortunately, lattice simulations can only determine the masses of hadronic ground states and low excited states for given spin and parity \cite{Montvay:1994cy}. Here we point out that recent lattice simulations of QCD at finite temperature in the range of temperatures $T<T_c \approx 160$ MeV \cite{Borsanyi:2010cj} are sensitive to the hadronic mass spectrum and permit to determine or, at least, constrain it beyond the experimentally established range. Earlier similar studies \cite{Karsch:2003vd} which, however, were either based on lattice simulations with unphysically high quark masses and focused on the baryon sector of the hadron mass spectrum, or were performed for lattice actions that resulted in significantly higher values of $T_c$ \cite{Huovinen:2009yb}. Other recent studies of high-mass resonance states were mainly aimed at their contributions to transport properties of QCD matter below the deconfinement temperature\cite{NoronhaHostler:2007jf,NoronhaHostler:2008ju,NoronhaHostler:2009cf}.
 
The so-called {\em interaction measure} (the QCD trace anomaly),
\be
\label{eq:I(T)}
I(T) = (\epsilon-3P)/T^4 ,
\ee
derived from the trace of the stress-energy tensor $T_\mu^\mu = \epsilon-3P$, is especially sensitive to the presence of massive states. To see why this is so, we calculate $I(T)$ for a thermal gas of non-interacting hadrons with mass spectrum $\rho(m)$:
\ba
\label{eq:I-HRG}
I(T) &=& \int_0^\infty dm\, \rho(m) \int\frac{d^3p}{(2\pi)^3E} ( E^2 - {\bf p}^2) e^{-E/T}
\nn \\
&=& \frac{1}{2\pi^2T^3} \int_0^\infty m^3 dm\, \rho(m)\, K_1(m/T) .
\ea
where we assumed that the hadrons are on mass shell: $E^2={\bf p}^2+m^2$. Because the trace anomaly vanishes in the conformal limit, light hadrons contribute little to $I(T)$. Heavy hadrons, on the other hand, contribute disproportionally due to the presence of the factor $m^3$ in the integrand in the last line of eq.~(\ref{eq:I-HRG}). In order to explore this quantitatively, we plot the integrand as a function of $m$ for several fixed temperatures in the range $T = 130-160$ MeV for an exponential mass spectrum of the form
\be
\label{eq:hagedorn1}
\rho(m) = c\, b\, e^{bm}
\ee
with $b=(252~{\rm MeV})^{-1}$ and $c=0.715$. For $T=160$ MeV (top curve) the integrand explores hadron masses much higher than for $T=130$ MeV (bottom curve). Since the mass range of well established hadron states only reaches up to approximately 1.5 GeV for non-strange mesons and 2 GeV for baryons, one expects that the interaction measure $I(T)$ is increasingly sensitive to experimentally unknown hadron states as the temperature exceeds 140 MeV.

\begin{figure}[!htb]
\centerline{\includegraphics[width=0.9\linewidth]{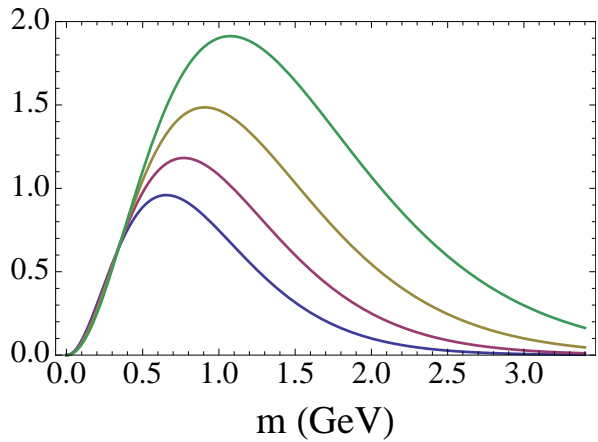}}
\caption{(Color online) The integrand of eq.~(\ref{eq:I-HRG}) in units of (1/GeV). The curves from bottom to top are for $T = 130,140,150,160$ MeV. Higher temperatures are seen to probe the hadron spectrum for increasingly larger hadron masses.}
\label{fig1}
\end{figure}

\begin{figure}[!htb]
\centerline{\includegraphics[width=0.9\linewidth]{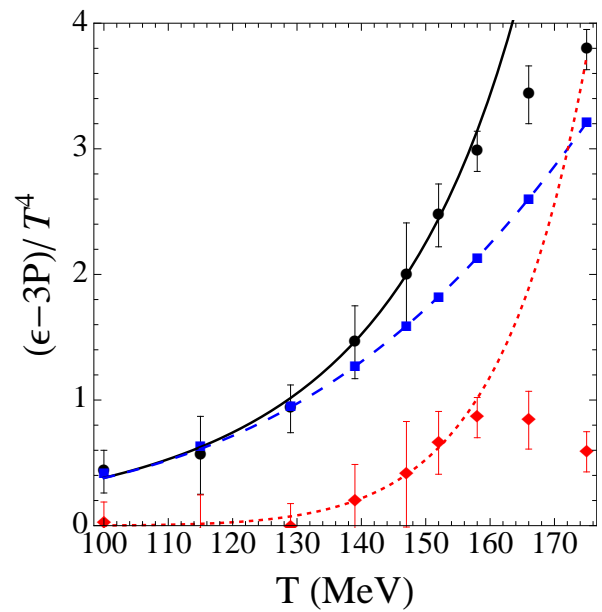}}
\caption{(Color online) QCD interaction measure $(\epsilon-3P)/T^4$ as a function of temperature $T$. The black dots (with error bars) show the results of a recent lattice calculation \cite{Borsanyi:2010cj}. The blue squares show the prediction of the hadron resonance gas model including all experimentally established states. The red diamonds represent the difference between the lattice results and the hadron resonance gas prediction. The lines show the predictions from eq.~(\ref{eq:I-HRG}) for an exponential hadron mass spectrum. The solid (black) curve includes all masses; the dashed (blue) curve only masses up to 1.7 GeV; and the dotted (red) curve only masses above 1.7 GeV. The parameters of the mass spectrum are given in the text.}
\label{fig2}
\end{figure}

This expectation is borne out by our results for the temperature dependence of the interaction measure. Figure~\ref{fig2} shows the results of a recent lattice calculation \cite{Borsanyi:2010cj} as black dots (with error bars). The blue squares show, in comparison, the prediction of the hadron resonance gas (HRG) model (\ref{eq:I-HRG}) including the experimentally established states (23 non-strange mesons, 7 strange mesons, 19 non-strange baryons, and 17 baryons with open strangeness), combining to a total of 795 
states~\cite{Eidelman:2004wy}. The red diamond symbols (with error bars) represent the difference between the lattice results and the HRG prediction. One notices that the HRG with the experimentally known mass spectrum provides an excellent description of the lattice results for temperatures up to $T\approx 130$ MeV, but then falls below the lattice results. 

In order to understand the deviation we show, as the dashed (blue) curve,  a prediction of the exponential mass spectrum (\ref{eq:hagedorn1}) with an upper cut-off at $m_c = 1.7$ GeV. The curve is seen to be in excellent agreement with the predictions of the HRG including only the experimentally known hadrons. The cut-off $m_c$ represents a reasonable compromise between the upper limits of the known meson and baryon spectra. When we remove the cut-off and integrate without a limit placed on the allowed hadronic mass, we obtain the solid (black) curve, which follows the lattice results for the interaction measure up to $T = 158$ MeV. The dotted (red) curve, which includes only masses \emph{above} 1.7 GeV, provides a good representation for the difference between the lattice results and the HRG curve. Again, this difference only seems to be describable by a noninteracting hadron gas model up to a temperature of $T = 158$ MeV.

The deviation of the HRG curve with an unlimited exponential mass spectrum from the lattice results for $T > 155$ MeV is not surprising, because this temperature approaches the quasi-critical temperature $T_c \approx 160-165$ MeV where deconfinement and chiral symmetry restoration sets in \cite{Borsanyi:2010bp}. The HRG cannot provide a description of these phenomena. 

\begin{figure}[!htb]
\centerline{\includegraphics[width=0.9\linewidth]{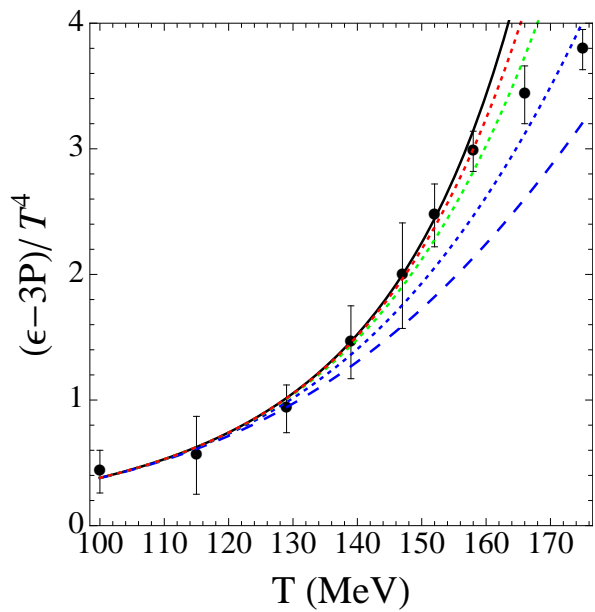}}
\caption{(Color online) QCD interaction measure $(\epsilon-3P)/T^4$ as function of temperature $T$. The black dots (with error bars) show the results of a recent lattice calculation \cite{Borsanyi:2010cj}. The lines show the predictions from eq.~(\ref{eq:I-HRG}) for an exponential hadron mass spectrum with different cut-off masses $m_c$. The solid (black) curve includes all masses; the dashed (blue) curve only masses up to 1.7 GeV.  The dotted curves show the results for mass cut-offs $m_c$ of $2.0$ (bottom, blue), $2.5$ (middle, green), and $3.0$ GeV (top, red), respectively.}
\label{fig3}
\end{figure}

In order to explore the sensitivity of the interaction measure to the mass cut-off in the hadron mass spectrum, we show in Fig.~\ref{fig3} the HRG predictions for the exponentially growing mass spectrum~\eqref{eq:hagedorn1} with different cut-offs. The solid (black) curve shows the prediction of an unlimited spectrum. The dashed (blue) curve shows the prediction for an upper cut-off of $m_c = 1.7$ GeV. The dotted curves show the results for upper mass cut-offs $m_c$ of $2.0$ (bottom, blue), $2.5$ (middle, green), and $3.0$ GeV (top, red), respectively. From the comparison of these curves with the lattice results (black dots with error bars), we see that at least a cut-off of $m_c = 2.5$ GeV is needed to fit the lattice results up to $T = 158$ MeV. It is also clear that a significantly higher precision of the lattice data would permit to probe the cut-off dependence to even higher values of $m_c$. However, Fig.~\ref{fig3} also clearly indicates that there is practically no sensitivity to masses above 3 GeV.

To explore the sensitivity of our results to the analytic form of the resonance gas spectrum, we have repeated our calculation for a mass spectrum of the form originally proposed by Hagedorn \cite{Hagedorn:1965st}
\be
\label{eq:hagedorn2}
\rho_{\rm H}(m) = \frac{A\,e^{m/T_{\rm H}}}{(m^2+m_0^2)^{5/4}}.
\ee
The parameters were fit by comparing with the hadron resonance gas model prediction. An equally good fit as that in Fig.~\ref{fig2} is obtained for the parameters $T_{\rm H} = 180$ MeV, $A = 0.63~{\rm GeV}^{3/2}$ and $m_0 = 0.5$ GeV. We will refer to this spectrum as the Hagedorn mass spectrum. The one difference is that in this case the cut-off mass must be chosen as $m_c = 1.9$ GeV. Calculating the interaction measure with an unbounded distribution, {\em i.~e.}, setting $m_c = \infty$, we obtain an identical prediction as in the case of Fig.~\ref{fig2} for the temperature ($T\approx 155$ MeV) where the Hagedorn mass spectrum begins to deviate from the lattice data. 

\begin{figure}[!htb]
\centerline{\includegraphics[width=0.9\linewidth]{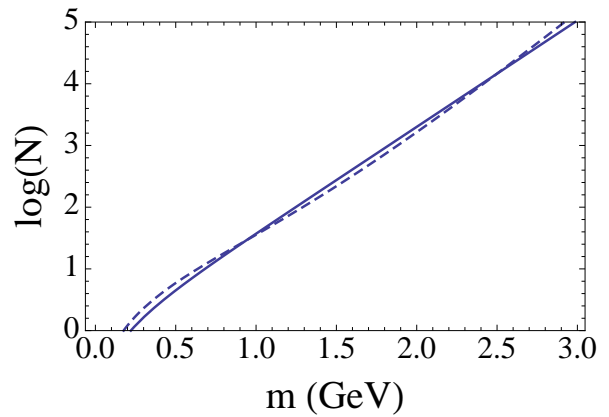}}
\caption{Number of hadron states below mass $m$ shown on a logarithmic scale. The solid curve is the prediction of the exponential mass spectrum (\ref{eq:hagedorn1}); the dashed line represents the prediction of the spectrum (\ref{eq:hagedorn2}) originally proposed by Hagedorn.}
\label{fig4}
\end{figure}

It may be surprising how two different parametrizations for the hadron mass spectrum can yield the same result for the temperature dependence of the interaction measure. This is due to two reasons: As pointed out above, in the relevant temperature range ($T < 155$ MeV) there seems to be very little sensitivity to resonances heavier than $3$ GeV.  Although the limiting temperature implied by the Hagedorn mass spectrum is considerably smaller (180 MeV compared with 252 MeV) than that associated with the exponential spectrum (\ref{eq:hagedorn1}), both predict nearly identical hadron mass distributions below 3 GeV, as shown in Fig.~\ref{fig4}. As a result, the predictions for the interaction measure from these two parametrizations differ only at very high temperatures, where they are no longer relevant to the lattice data.

We conclude that recent lattice simulations of QCD with $2+1$ dynamical quark flavors and physical quark masses already provide evidence for the existence of hadron states beyond those experimentally known. The interaction measure (\ref{eq:I(T)}), which is especially sensitive to high-mass hadron states, is in good agreement with an exponential mass spectrum of hadrons up to $T=158$ MeV, but a mass spectrum of the Hagedorn type can also provide a good fit.  Increasingly precise future lattice simulations will enable a precision determination of the hadron mass spectrum beyond the range where individual hadron states can be experimentally resolved. However, our analysis demonstrates that the interaction measure at temperatures below $T_c$ is not sensitive to hadron masses above 3 GeV.

To obtain additional information about the hadronic mass spectrum, we propose the study of the variation of the interaction measure with baryon chemical potential, strangeness chemical potential and/or quark flavor chemical potentials~\cite{Koch:2005vg}. A study in this vein was carried out in Ref.~\cite{Chatterjee:2009km}, who used a hybrid model, {\em i.~e.}, a combination of known resonances below a mass cut-off and a Hagedorn spectrum above the cut-off, to explore the sensitivity of the energy density and the pressure to the parametrization chosen. However, to the best of our knowledge no study of the dependence of the interaction measure on quark chemical potentials has been been carried out. While we do not present any quantitative predictions in this Letter due to the scarcity of lattice data on this topic, the results of such a study may be easily discerned: The introduction of a baryon chemical potential will require one to decompose the hadron distributions in Eq.~\eqref{eq:I-HRG} into a mesonic and baryonic part with the Boltzmann distribution of the baryonic piece modified to include the baryon chemical potential, i.~e., $\exp(-E/T ) \rightarrow \exp( - [E - \mu]/T )$ with the implied opposite sign for anti-baryons. Derivatives of the interaction measure with respect to $\mu$ will allow for an estimation of the portion of these unknown states which lie in the baryon spectrum. Even more detailed analyses may be performed by the introduction of separate chemical potentials for each quark flavor.

\begin{acknowledgments}  
This work was supported in part by grants from the U.S. Department of Energy (DE-FG02-05ER41367 and DE-FG02-01ER41190) and the BMBF. The research reported here was initiated during the HESI10 Workshop at the Yukawa Institute of Kyoto University. We thank A.~Sch\"afer and S.~A.~Bass for helpful discussions and C.~Greiner for comments on an earlier version of the manuscript.
\end{acknowledgments}

\end{document}